\documentclass{river-journal}
\pdfoutput=1
\usepackage{newtxtext,newtxmath}
\usepackage{graphicx}
\usepackage{graphics}
\usepackage{subfigure}
\usepackage{cite}
\usepackage{url}
\usepackage{amsmath}
\usepackage{multirow}
\usepackage{verbatim}

\begin{document}
	\date{\phantom{x}}
\begin{opening}
\title{Composite Metrics for Network Security Analysis}
\author{Simon Yusuf Enoch, Jin B. Hong, Mengmeng Ge and Dong Seong Kim}
\institute{Department of Computer Science and Software Engineering, University of Canterbury, Private Bag 4800, Christchurch, New Zealand.  \texttt{sey19@uclive.ac.nz, \{jho102, mge43\}@uclive.ac.nz and dongseong.kim@canterbury.ac.nz}}
\end{opening}

\runningtitle{Composite Metrics for Network Security Analysis}
\runningauthor{S. Y. Enoch et al.}

\subsection*{Abstract}
Security metrics present the security level of a system or a network in both qualitative and quantitative ways. In general, security metrics are used to assess the security level of a system and to achieve security goals. There are a lot of security metrics for security analysis, but there is no systematic classification of security metrics that is based on network reachability information. To address this, we propose a systematic classification of existing security metrics based on network reachability information. Mainly, we classify the security metrics into host-based and network-based metrics. The host-based metrics are classified into metrics ``without probability" and ``with probability", while the network based metrics are classified into ``path-based" and ``non-path based". Finally, we present and describe an approach to develop composite security metrics and it's calculations using a Hierarchical Attack Representation Model (HARM) via an example network. Our novel classification of security metrics provides a new methodology to assess the security of a system.


\section{Introduction}
Researchers from research institutions, governments and industries have been working on developing and distributing security metrics. For instance, the Center for Internet Security (CIS) \cite{CIS:SecurityMetric2010} proposed and categorised security metrics into management, technical and operational metrics. The National Institute of Standards and Technology (NIST) \cite{NIST:NIST2007} proposed nine security metrics into implementation, effectiveness/efficiency and impact. Others such as Idika and Bhargava \cite{Idika:Extending} proposed and classified security metrics into decision, assistive   and so on. Most of these efforts to categorise and classify security metrics are based on the target audience and personal intuitions. Therefore, it is important to develop a systematic classification of security metrics that is based on network reachability information.
There are a number of security metrics which are used for network security assessment \cite{Idika:Extending, Phillipsi:VulAnalysis, Pamula:WA,Wang:AR,Wang:PS}. But none of them are capable of representing the overall security level of the network \cite{Krautsevich2011}. Thus it is important we combine different security metrics to present and analyse the diverse facet of the security posture.

In this paper, we classify the existing security metrics based on network reachability information, and describe an approach to develop new security metrics by combining the existing security metrics. Our novel classification provides a new methodology to assess the security of a system. It also provides insight as to how and when a security metric should be used. 
The main contributions of this paper are:
\begin{itemize}
	\item to classify existing cyber security metrics;
	\item to perform security analysis using the existing security metrics;
	\item to describe an approach to developing composite security metrics; and
	\item to formally define the composite security metrics.
\end{itemize}

The rest of the paper is organised as follows. Section~\ref{RelWork} introduces related work on existing classification of security metrics. In Section~\ref{MetricsClass}, we present a novel classification of the existing security metrics. In Section \ref{AnalysisEgNet}, we describe and analyse the security of an example network using existing security metrics. In Section \ref{NewMetrics}, we present our new composite security metrics with examples. And finally, we conclude the paper and outline the future work in Section~\ref{conclusions}.

\section{Related Work}
\label{RelWork} 
There are a few research on the classification of security metrics. Most classification methods are based on organisation's point of view \cite{Savola:Taxonomy2007}. For instance, Savola \cite{Savola:metricTaxonomy2007} proposed three categories of security metrics; namely, (i) business-level security metrics, (ii) metrics for information security management (ISM) in organisations, and (iii) dependability and trust metrics for products, systems and services. The business-level security metrics are business goals directed and are used for cost-benefit security analysis in organisations. The information security management metrics are used to evaluate the ISM security controls, plans and policies, and are divided into three subcategories (i.e., management, operational and information system technical security metrics). The dependability and trust metrics are used to assess the organisation's trust, relationships and dependability issues \cite{Avizienis:Dependability2004}. In general, this classification only addresses the security needs of companies that produce information and telecommunication technology products, systems or services.

Vaughn \textit{et al.} in \cite{Vaughn:SecMet2002} presented two categories of security metrics (organisational security metrics and metrics for technical target assessment). The organisational security metrics assess the organisation's security assurance status (the metrics in this category include security effectiveness, operational readiness for security incidents and information assurance program development metric). The metrics for technical target assessment are used to assess the security capabilities of a technical system (it is further divided into metrics for strength assessment and  metrics for weakness assessment \cite{Vaughn:SecMet2002}). This classification is tailored towards an organisation's needs. 

Pendleton \textit{et al.} \cite{Pendleton:Survey2016} classified security metrics into four categories, namely: metrics for measuring the system vulnerabilities, metrics for measuring the defences, metrics for measuring the threats, and metrics for measuring the situations. 
The metrics for measuring vulnerabilities are intended to quantify the enterprise and computer systems vulnerabilities through their user's password, software vulnerabilities, and the vulnerabilities of the cryptographic keys they use. The metrics for measuring defences is aimed to quantify the countermeasure deployed in an enterprise via the effectiveness of blacklisting, the ability of attack detection, the effectiveness of software diversification, and the overall effectiveness of these countermeasures. The metrics for measuring threats are aimed to assess the threats against an enterprise through the threat of zero-day attacks, the power of individual attacks and the sophistication of obfuscation. And the metrics for measuring the situations aims to assess situations via security investments, security states and security incidents. This classification is centred on the perspective between attackers and defenders in enterprise systems. Other classifications provided by industries such as the NIST \cite{NIST:NIST2007}, the CIS \cite{CIS:SecurityMetric2010} and the Workshop on Information Security System Scoring and Ranking are exclusively geared towards cyber defence administrations and operations  \cite{Pendleton:Survey2016}. 

To the best of our knowledge, there is no previous work on the classification of security metrics based on the network reachability information. Here, we focus on classifying existing security metrics based on network reachability information and propose an approach to develop new set of cyber security metrics by combining the existing metrics.


\section{Classification of Security Metrics}
\label{MetricsClass} 
\begin{figure}[hbt!]
	\centerline{\includegraphics[width=5in]{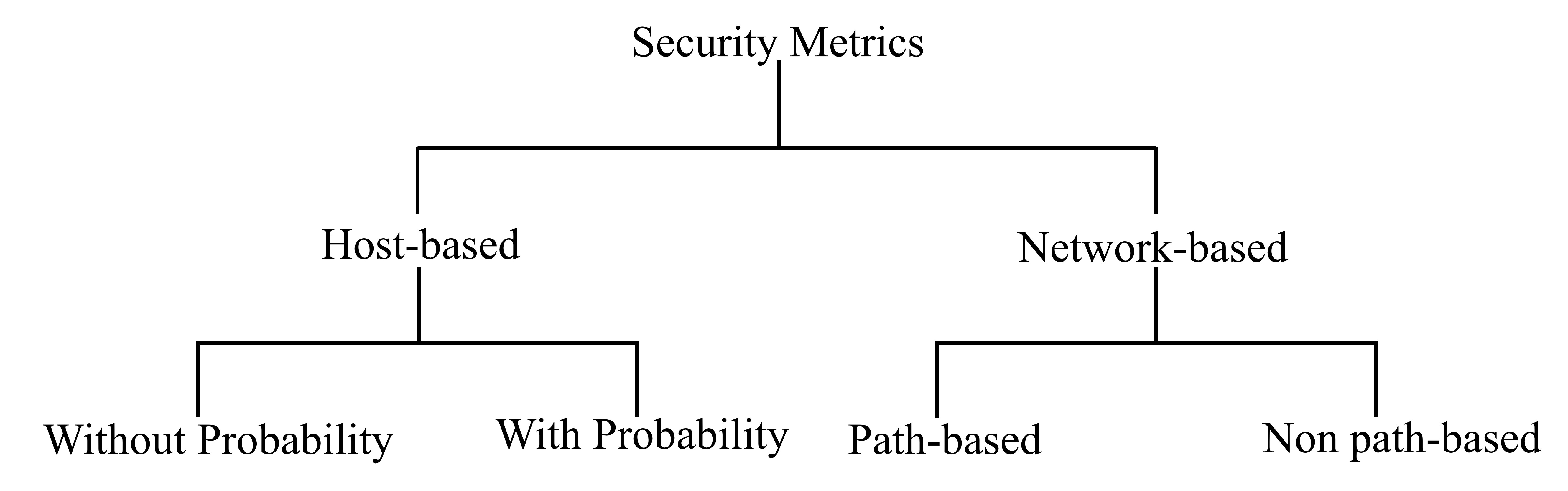}}
	\caption[]{Classification of Security Metrics.}%
	\label{fig_MetricsClass}
\end{figure}

Based on network reachability information, we mainly classify security metrics into two types: host-level metrics and network-level metrics, as shown in Figure~\ref{fig_MetricsClass}. 

The host-level metrics do not use any network level information (e.g., reachability, protocols, \textit{etc}) whereas the network-level metrics take into account network structure, protocol and reachability information to quantify the security of a system. We describe the host-level metrics in Section \ref{hostlevel} and the network-level metrics in Section \ref{networklevel}, respectively. 

\subsection{ Host-based Security Metrics}
\label{hostlevel}
The host-level metrics are used to quantify the security level of individual hosts in a network. We further classify the host-level metrics into two types: ``without probability" and ``with probability". The reasons for this classification are: (i) sometimes it is infeasible to find a probability value for an attack, and (ii) some analysis and optimisation can be done with or without probability assignments as described in \cite{Roy:OptimumC2012}. 

\subsubsection{Metrics without probability values}
We summarise the metrics ``without probability'' in Table~\ref{tablewithout}. Examples of metrics without probability values are attack impact, attack cost, structural important measured \cite{Roy:ACT},  mincut analysis \cite{Roy:ACT}, mean-time-to-compromise (MTTC)\cite{Leversage:MTTC,Ge:IOT_mttc2015}, mean-time-to-recovery (MTTR) \cite{Jaquith:ReplacingFear}, Mean-Time-to-First-Failure (MTFF) \cite{Sallhammar:Stochastic}, Mean-Time-to-Breach (MTTB)\cite{Jonsson:MTTB}, The return on investment \cite{Cremonini:ROA2005}, The return on attack \cite{Cremonini:ROA2005}, \textit{etc}. 

\begin{table}[hbt!]
	\caption[]{Description of Metrics without Probability Values.}
	\label{tablewithout}
	\begin{tabular}{ll} \hline
		Metrics                                                    &  Description                              \\ \hline
		Attack Cost  \cite{Roy:ACT} & is the cost spent by an attacker to successfully\\& exploit a vulnerability (i.e., security weakness)\\& on a host. 
		\\ \hline 
		\multirow{1}{*}{Attack Impact \cite{Hong:Thesis}}& is the quantitative measure of the potential\\& harm caused by an attacker to exploit a\\& vulnerability.\\ \hline
		\multirow{2}{*}{Mean-time-to-Compromise (MTTC)\cite{Leversage:MTTC,Ge:IOT_mttc2015}} & is used to measure how quickly a network can \\&be penetrated. This type of metrics produces \\&time values as end results.\\
		\hline
		\multirow{1}{*}{Structural Important Measure  \cite{Roy:ACT}}& is used to qualitatively determine the most \\&critical event (attack, detection or mitigation)\\& in a graphical attack model. This metric is \\&useful when the probability of event such as \\&attack, detection or mitigation are unknown.\\ \hline
		\multirow{2}{*}{ Mean-Time-to-Recovery (MTTR)\cite{Jaquith:ReplacingFear}} & is used to assess the effectiveness of a network \\&to recovery from an attack incidents. It is \\&defined as the average amount of time required\\& to restore a system out of attack state. The \\&shorter the time, the less impact is the attack \\&on the overall performance of the network.\\
		\hline
		\multirow{2}{*}{The Return on Attack \cite{Cremonini:ROA2005}} & is defined as the gain the attacker expects from\\& successful attack over the losses he sustains due\\& to the countermeasure deployed by his target. \\&This security metric is from the attacker \\&perspective and it used by organisations to \\&evaluate the effectiveness of a countermeasure\\& in discouraging a certain type of intrusion \\&attempts \cite{Cremonini:ROA2005}.  \\
		\hline
	\end{tabular}
\end{table}

\subsubsection{Metrics with probability values}
Conversely, the security metrics with probability include probability security metric \cite{Wang:PS}, Common Vulnerability Scoring System (CVSS) metrics \cite{CVSS:2016} \textit{etc}. An attack graph (AG) is an acyclic directed graph to represent all possible ways for an attacker to reach a target vulnerability. Wang \textit{et al.} \cite{Wang:PS} proposed an AG-based security metric that incorporates the likelihood of potential multi-step attacks combining multiple vulnerabilities in order to reach the attack goal. We summarise the metrics with probability in Table \ref{tableProb}.

\begin{table}[hbt!]
	\caption[]{Description of Metrics with Probability Values.}
	\label{tableProb}
\begin{tabular}{ll} \hline
	Metrics                                                    &  Description                              \\ \hline
	Probability of vulnerability exploited \cite{Edge:Thesis} & \begin{tabular}[c]{@{}l@{}}is used to assess the likelihood of an attacker \\exploiting a specific vulnerability on a host. This \\takes into account the severity of the host\\ vulnerability.\end{tabular} \\ \hline
	Probability of attack detection  \cite{Roy:ACT}  & \begin{tabular}[c]{@{}l@{}}is used to assess the likelihood of a \\countermeasure to successfully identify the \\event of an attack on a target.\end{tabular} \\ \hline
	Probability of host compromised \cite{Homer:Metric2013}& \begin{tabular}[c]{@{}l@{}}is used to assess the likelihood of an attacker to \\successfully compromise a target \end{tabular} \\ \hline
	CVSS \cite{CVSS:Guide,CVSS:2016}& \begin{tabular}[c]{@{}l@{}}is an industry standard used to assess the severity\\ of computer vulnerabilities. Details of the CVSS \\probability is provided in \cite{Ou:RiskAssessment}. \end{tabular} \\ \hline
	\end{tabular}
\end{table}


\subsection{Network-based Security Metrics}
\label{networklevel}
This category of metrics uses the structure of a network to aggregate the security property of the network. We further classify these metrics into two types: path based and non-path based metrics (according to the use of path information). 
\begin{table}[hbt!]
	\caption[]{Description of Path based Metrics.}
	\label{PathAnalysis}
	\begin{tabular}{ll} \hline
		Metrics                                                    &  Description                              \\ \hline
		\multirow{2}{*}{Attack Shortest Path \cite{Phillipsi:VulAnalysis,Ortalo:OperationSec}}& is the smallest distance from the attacker to the \\ &target. This metric represents the minimum \\ &number of hosts an attacker will use to \\ &compromise the target host.\\ \hline
		\multirow{2}{*}{Number of Attack Paths \cite{Ortalo:OperationSec}} & is the total number of ways an attacker can com-\\ &promise the target. The higher the number, the \\ &less secure the network.\\ \hline
		\multirow{2}{*}{Mean of Attack Path Lengths \cite{Li:ExploitGraph}} & is the average of all path lengths. It gives the \\ &expected effort that an attacker may use to \\ &breach a network policy. \\ \hline
		\multirow{2}{*}{Normalised Mean of Path Lengths \cite{Idika:Extending}} & This metric represents the expected number of \\ &exploits an attacker should execute in order to \\ &reach the target.\\ \hline
		Standard Deviation of Path Lengths\cite{Idika:Extending} & is used to determine the attack paths of interest.\\&A path length that is two standard deviations \\&below the mean of path length metric is \\&considered the attack paths of interest and can \\&be recommended to the network administrator for\\& monitoring and consequently for patching \cite{Idika:Extending}.\\ \hline
		Mode of Path Lengths \cite{Idika:Extending} & is the attack path length that occurs most \\&frequently. The Mode of Path Lengths metric \\&suggests a likely amount of effort an attacker\\& may encounter.\\ \hline
		Median of Path Lengths \cite{Idika:Extending} & this metric is used by network administrator to \\&determine how close is an attack path length\\& to the value of the median path length (i.e. path\\& length that is at the middle of all the path\\& length values). The values that falls below the\\& median are monitored and considered for\\& network hardening \cite{Idika:Extending}.\\ \hline
		\multirow{2}{*}{Attack Resistance Metric \cite{Wang:AR}} & is use to assess the resistance of a network confi-\\ &guration based on the composition of measures of \\ &individual exploits. It is also use for assessing and \\&comparing the security of different network \\&configurations \cite{Wang:AR}. \\
		\hline
	\end{tabular}
\end{table}

\subsubsection{Non-Path Based Metrics}
In non-path based metrics, the structure and attributes of a network are not considered; instead, the security of a network is quantified regardless of the network structure. One example of this type of metrics is Network Compromise Percentage (NCP) metric \cite{Lippmann:NCP}. The NCP metric is defined in Table~\ref{nonpath}. This metric indicates the percentage of network assets an attacker can compromise. The aim of the NCP metric is to minimise this percentage. Another example is a set of vulnerabilities that allows an attacker to use them as entry points to a network. For instance, web-services running on a host could be the very first targets for an attacker to compromise. The weakest adversary (WA) metric is also a network based metric that is use to assess the security of a network. In the WA metric, a network configuration that is vulnerable to a stronger set of attribute is define as more secure than a network configuration that is vulnerable to a weaker set of initial attacker attributes \cite{Pamula:WA}.

\begin{table}[hbt!]
	\caption[]{Description of Non-Path Based Metrics.}
	\label{nonpath}
	\begin{tabular}{ll} \hline
		Metrics                                                    &  Description                              \\ \hline
		Network Compromise Percentage \cite{Lippmann:NCP} & \begin{tabular}[c]{@{}l@{}}is the metric that quantifies the \\percentage of hosts on the network\\ on which an attacker can obtain an\\ user or administration level privilege. \end{tabular} \\  \hline
		Weakest Adversary \cite{Pamula:WA} & \begin{tabular}[c]{@{}l@{}} is used to assess the security strength of a \\network  in terms of the weakest part of \\the network that an attacker can \\successfully penetrate.\end{tabular} \\
		\hline
		Vulnerable Host Percentage \cite{Kott:SecurityMetic2014} & \begin{tabular}[c]{@{}l@{}} is used to assess the overall security of \\a network. This metric quantifies the \\percentage of hosts with vulnerability on a \\network. The higher the metric value, \\the less is the security level of the network.\end{tabular} \\
			\hline
	\end{tabular}
\end{table}

\subsubsection{Path Based Metrics}
Path based metrics use the reachability information of a network (for example, reachability between hosts, shortest path from a host \textit{X} to a host \textit{Y}, and so on) to quantify the security level of the network. We summarise some of these metrics in Table~\ref{PathAnalysis}, which include Shortest Path (SP) metrics \cite{Phillipsi:VulAnalysis}, Number of Paths (NP) metrics \cite{Ortalo:OperationSec}, Mean of Path Length (MPL) metrics \cite{Li:ExploitGraph}, Normalised Mean of Path Lengths (NMPL) Metrics \cite{Idika:Extending}, Standard Deviation of Paths Lengths (SDPL) Metrics \cite{Idika:Extending}, Mode of Path Lengths (MoPL) Metrics \cite{Idika:Extending} and Median of Path Lengths (MePL) Metrics \cite{Idika:Extending}.

\section{Network Configurations and System Model}
\label{AnalysisEgNet}
The example network is shown in Figure \ref{fig_Example}. The network consists of two firewalls with an attacker located outside the network. Here, the firewall 1 is use to allow secure connections from the Internet to the hosts in the network while firewall 2 is use to allow secure connections to the database (i.e., $h_7$).
We assume the goal of the attacker is to compromise the database. We denote hosts in the network as $h_i$, where $i=1,2,3...,n$ (a unique identifier for each host in the network). Table \ref{tb_routing_net1} shows the firewall rules used for the example network. For simplicity, we selected only one vulnerabilities for each host in the network from the Common Vulnerabilities and Exposures (CVE) \cite{CVSS:2016} which we list in Table \ref{ListVul}. 

We use the example network and existing security metrics to perform security assessment via the Hierarchical Attack Representation Model (HARM) \cite{Hong:HARM_MTD2015}. 
We describe the HARM and assumptions of the example network in Section \ref{Harm} and Section \ref{assumptions}, respectively.

The example network has a finite set of hosts $H$ and a finite set of vulnerabilities $V$. The following notations are used for the security assessment.

\begin{table}[hbt!]
	\caption[]{Notations for the security assessment.}
	\label{tblNotation}
	\begin{tabular}{ll}
		\hline
		\it{Notation}                                                      &  \it{Meaning}                                \\ \hline
		$AP$                                                            & is all possible paths from an attacker to a target                   \\
		\begin{tabular}[c]{@{}l@{}}$ap$\end{tabular} & is an attack path which includes a sequence of hosts   \\ 
		\begin{tabular}[c]{@{}l@{}}$f$ \end{tabular} & is a function that identifies the length of the attack path that occurs most\\& frequently\\ 
		$ac_{h}$ & is the minimum cost spent by an attacker who successfully compromises\\& the host $h$ \\ 
		$aim_{h}$ & is the maximum potential loss caused by an attacker who successfully \\&compromises the host $h$ \\ 
		$pr_{h}$ & is the probability of an attacker to successfully compromise the host $h$ \\ 
		$ac_{ap}$ & is the minimum cost spent by an attacker who successfully compromises\\& an $ap$ \\ 
		$aim_{ap}$ & is the maximum potential loss caused by an attacker who successfully \\&compromises an $ap$ \\ 
		$pr_{ap}$ & is the probability of an attacker to successfully compromise an $ap$ \\ 
		$ap_{_{ex}}$ & is the attack path that an attacker is attempting to exploit $e_x$\\
		$as_h$ & is the asset value associated with a host $h$\\ 
		$s_v$ & is the set of vulnerable hosts\\ 
		\hline
	\end{tabular}
\end{table}

\begin{itemize}
	\item A graphical security model - HARM denoted as $GSM$
	\item Each host $h \in H$ has a name $h_{name}$, a vulnerability $v \in V$ and a set of security metrics $h_{metrics}$$\subseteq$\{$p_h$, $iam_h$, $ac_h$, $mttc_h$, $as_h$\}.
	\item Each vulnerability $v \in V$ has a name $v_{name}$.
	\item Each attack path $ap \in AP$ has an index $ap_{index}$.
	
\end{itemize}

\subsection{The HARM}
\label{Harm}
We use the HARM to analyse the network security. The HARM is a two-layer model in which the upper layer (AG) represents the network reachability information and the lower layer (AT) represents the vulnerability information. 

We defined the AT \cite{Ge:iot2017} for HARM as a 5-tuple $at=(A, B, c, g, root)$. Here, $A$ is a set of components which are the leaves of $at$ and $B$ is a set of gates which are the inner nodes of $at$. We require $A{\cap}B=\emptyset$ and $root{\in}A{\cup}B$. Function $c$: $B{\rightarrow}\mathcal{P}(A{\cup}B)$ describes the children of each inner node in $at$ (we assume there are no cycles). Function $g$: $B{\rightarrow}\{AND, OR\}$ describes the type of each gate. The representation of the AT $at_h$ associated to the host $h$$\in$$H$ is given as $at_h$: $A{\subseteq}h_{vuls}$ (where $vuls$ is the host vulnerability). This means that the vulnerabilities of a node are combined using $AND$ and $OR$ gates.

We defined the AG for HARM \cite{Ge:iot2017} as a directed graph $ag=(N,E)$ where $N$ is a finite set of components and $E{\subseteq}N{\times}N$ is a set of edges between components.

\begin{figure}[hbt!]
	\centering
	\subfigure[Example Network.]{
		\label{egNet}
		\includegraphics[width=0.9\textwidth]{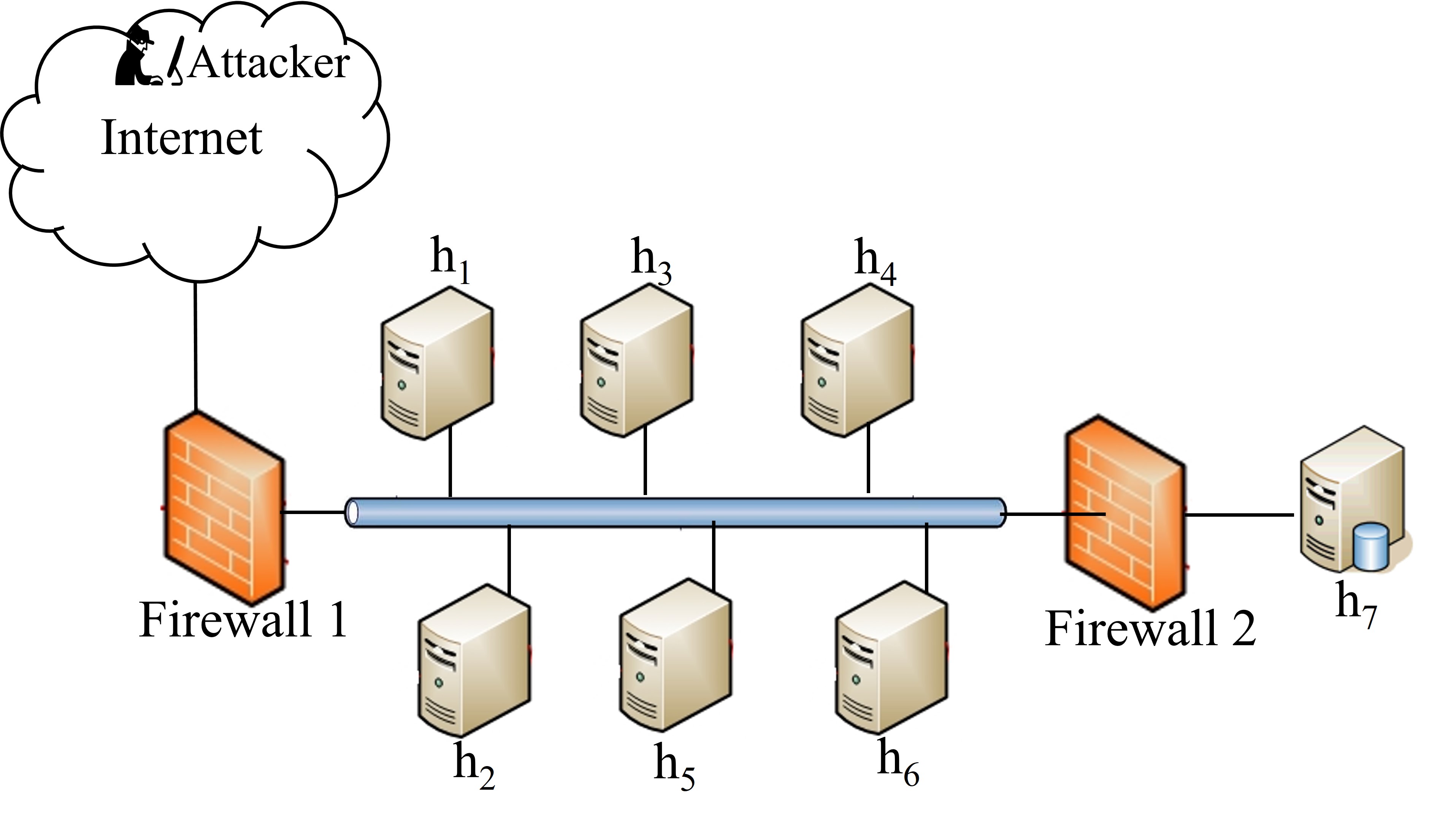}
	}
	\subfigure[HARM of the Example Network.]{
		\includegraphics[width=0.60\textwidth]{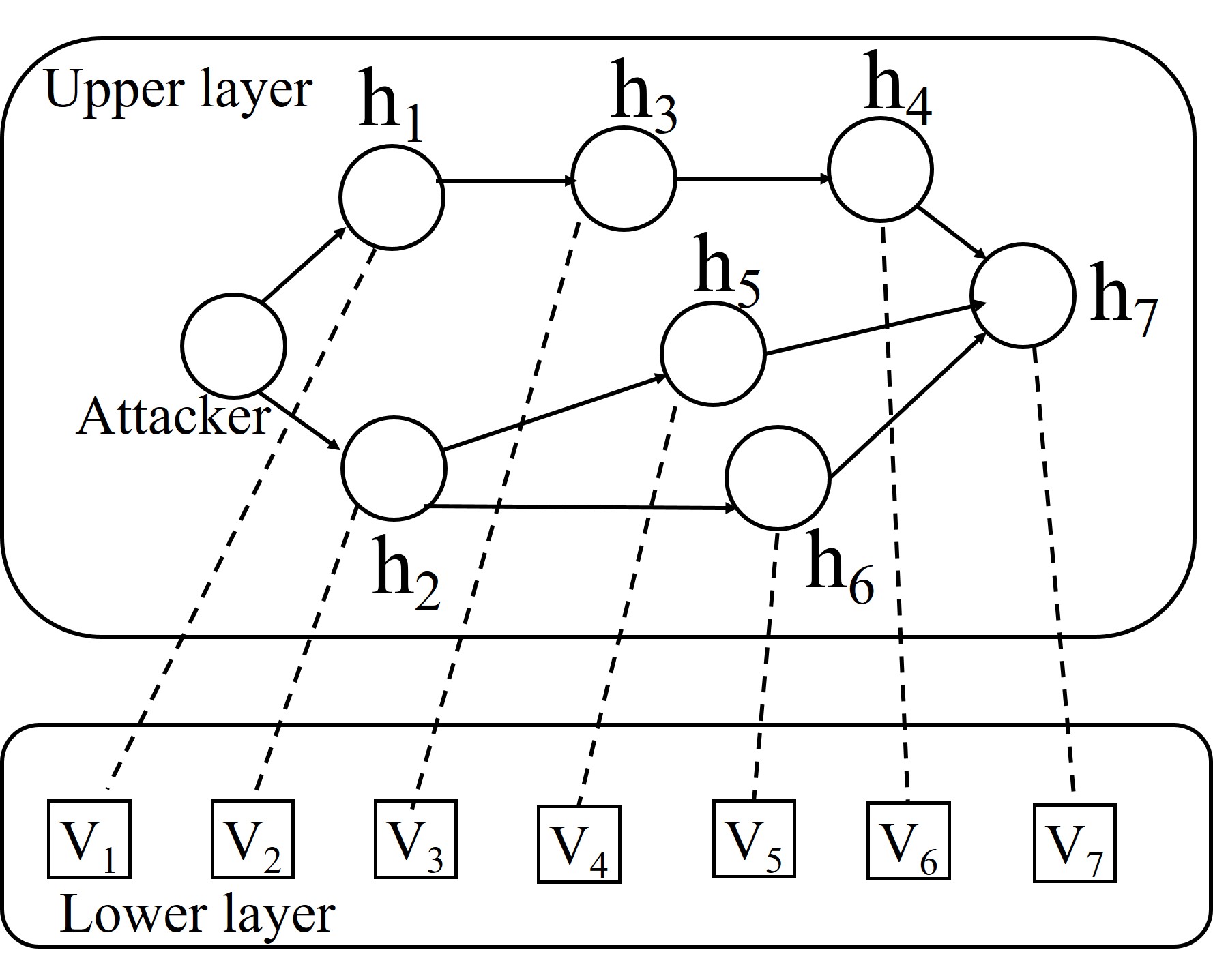}
		\label{egHARM}
	}
	\caption{An Example Network and the HARM}
	\label{fig_Example}
\end{figure}

The HARM of the example network is shown in Figure \ref{egHARM}. Other graphical security models such as those suggested by Noel and Jajodia \cite{Noel:SecurityRisk} and Ou \textit{et al.}\cite{Ou:AGgeneration} can also be used.


	\begin{table}[hbt!]
		\centering
		\caption[]{Example network: firewall rules.}
		\label{tb_routing_net1}
		\begin{tabular}{l|l}
			\hline
			Host &accept traffic from  \\ \hline 
			$h_1$  & Internet  \\  
			$h_2$  & Internet  \\  
			$h_3$  & $h_1$  \\  
			$h_4$ & $h_3$ \\  
			$h_5$ & $h_2$   \\ 
			$h_6$ & $h_2$\\ 
			$h_7$ & $h_4$, $h_5$, $h_6$  \\
			\hline
		\end{tabular}
	\end{table}

\subsection{Assumptions for the example network}
\label{assumptions}
We make the following assumptions for the example network:
\begin{itemize}
	\item An attacker knows the (or has knowledge of) reachability information from the attacker to the target (that is $h_7$).
	\item Each host has only one vulnerability but more vulnerabilities can be modelled as in the work~\cite{Hong:HARM,Hong:Thesis}.
	\item Exploiting a vulnerability grants the attacker the root privilege of the host.
	\item The attacker uses vulnerability scanners such as Nessus \cite{Deraison:scanner}, Nmap \cite{Nmap}, \textit{etc} to discover all the network vulnerabilities.
\end{itemize}

\subsection{Security Analysis of the Example Network}
\label{SecAnalysis}
We use existing security metrics to assess the security of the example network. For simplicity, we selected a few vulnerabilities from the Common Vulnerabilities and Exposures (CVE) \cite{CVSS:2016} which we list in Table \ref{ListVul}. 
 
\begin{table}[hbt!]
	\caption[]{List of Vulnerabilities.}
	\label{ListVul}
	\begin{tabular}{cccccccc}
		\hline
		\it{h name} &\it{v name} & \it{CVE-ID} & \it{CVSS BS} & $pr_h$ & $aim_h$ & $ac_h$  & $as_h$\\ \hline 
		$h_1$ &$v_1$ & CVE-2016-2386 & 7.5 & 0.75 & 7 & 8  & 40\\  
		$h_2$ &$v_2$ & CVE-2016-2040 & 3.5 & 0.35 & 4 & 4.2 & 21 \\  
		$h_3$ &$v_3$ & CVE-2016-0059 & 4.3 & 0.43 & 5 & 5 2 & 25\\  
		$h_4$ &$v_4$ & CVE-2015-7974 & 2.1 & 0.21 & 3 & 3.5 & 17.5\\ 
		$h_5$ &$v_5$ & CVE-2015-2542 & 9.3 & 0.93 & 9 & 9.2  & 46\\ 
		$h_6$ &$v_6$ & CVE-2014-2706 & 7.1 & 0.71 & 6.5 & 7.5  & 37.5\\ 
		$h_7$ &$v_7$ & CVE-2013-2035 & 4.4 & 0.44 & 4.3 & 5.5  & 27.5\\	
		\hline
	\end{tabular}
\end{table}

In Table \ref{ListVul}, the host-based metrics ``without probability" values; attack cost and attack impact have metric value of 5.50 and 4.30 for target host $h_7$, respectively. These metrics present the minimum cost and the potential loss for the attacker to successfully compromise a host $h_7$, respectively. The probability of attack success metric (i.e., a metric ``with probability") is -- 0.44. This metric presents the probability that an attacker will successfully exploit the host $h_7$. The lower the metric value, is the lower the chances that the attacker will succeed in exploiting the target host. 

To calculate the network base metrics, we consider a set of all attack paths $AP$ (i.e. $ap_1 = (h_1, h_3, h_4, h_7)$, $ap_2 = (h_2, h_5, h_7)$, and  $ap_3 = (h_2, h_6, h_7)$) for a given target, $h_7$. We compute the network based metrics in Table \ref{tableMetricPath} and Table \ref{tblNonPathBaseMetrics}. 

\begin{table}[hbt!]
	\caption[]{Metrics Values for ``Path based Metrics".}
	\label{tableMetricPath}
	\begin{tabular}{llc}
		\hline
		\it{Metrics Name}                                                      &  \it{Formulae}                                 & \it{Value} \\ \hline
		Shortest Attack Path                      & $SP(GSM)=\mathop {min}\limits_{ap \in AP}{|ap|} $                & 3.00              \\ \hline
		Number of Attack Paths                    & $NP(GSM)=|AP|$                & 3.00              \\ \hline
		Mean of Attack Path Lengths               & $MPL(GSM) = \frac{\sum\limits_{ap\in AP} {|ap|}} {NP(GSM)}$                & 3.30              \\ \hline
		SDPL  & $SDPL(GSM)=\sqrt{ \frac{\sum\limits_{ap\in AP}( {|ap|}-MPL(GSM))^2}{NP(GSM)}}$                & 0.47              \\ \hline
		Mode of Path Lengths               & $MoPL(GSM)=\mathop{f}\limits_{ap\in AP}{(|ap|)} $               & 3.00              \\ \hline
		Attack Resistance                 & $R(e_i) = \begin{cases}
		r(e_i)+R(e_j) & \text{conjuctive}\\ \hline
		r(e_i) + \frac{1}{R(e_k)^{-1} + R(e_l)^{-1}}  & \text{disjunctive} 
		\end{cases}$         & 8.81             \\ 
		\hline
	\end{tabular}
\end{table}


In Table \ref{tableMetricPath}, the value of the shortest path metric is 3. Based on this metric, an administrator can prioritise the network hardening measure by patching vulnerabilities along the shortest path --- in this case, it is the attack path $ap_2$ and $ap_3$. The number of paths (NP) metric which also yield the value 3.00 indicates the security strength of the network. In the NP metric, the higher the paths number is the lower is the security level. 
The mean of paths length yield 3.30. This security metrics show the overall network security level. In the mean of path lengths metric, the HARM with higher metric value is recorded as less secure.
The standard deviation of path lengths is 0.47. According to this metric, the path length that is two standard deviations below the mean of path lengths metric is considered to be the attacker's path of interest and regarded as vulnerabilities in hosts along the path are recommended for patching. In this case the  $ap_2$ and $ap_3$ are both two standard deviation below the MPL metric (their two standard deviation value is - 0.64 for both metrics).

To compute the attack resistance metric, two basic operators (disjunctive and conjunctive) described in Wang \textit{et al.} \cite{Wang:AR} are used. We compute the attack resistance metric based on the equation provided by Idika \cite{Idika:Thesis}. In the equation, the function $r$ represents the difficulty associated with an exploit $e_m$. $R$ represent the cumulative resistance of an exploit $e_m$ by taking into account all resistance values for ancestors of $e_m$. We use each host vulnerability value as the exploit value. In our calculation, the attack resistance value is 8.81. This metric value indicates the network security level and the ability of the network configuration to resist attack.




\begin{table}[hbt!]
	\caption[]{Metrics Values for ``non-path based metrics".}
	\label{tblNonPathBaseMetrics}
	\begin{tabular}{llc}
		\hline
		\it{Metrics Name}                                                      &  \it{Formulae}                                 & \it{Value} \\ \hline
		Network Compromise Percentage                                                      & $NCP(GSM)=100\times \frac{\sum\limits_{h\in ap_{_{ex}}}as_h}{\sum\limits_{h\in AP}as_h},  ap_{_{ex}}\in AP$             & 51.23\%  \\
		\hline
		Vulnerable Host Percentage                                                      & $VHP(GSM)=100\times \frac{\sum\limits_{h\in s_v}h}{\sum\limits_{h\in AP}h}, s_v\in AP $             & 100\%  \\	\hline
	\end{tabular}
\end{table}

In Table \ref{tblNonPathBaseMetrics}, we compute the NCP metric. The NCP security metric is for an AG that is not target oriented. In the NCP computation, we assume the attacker is attempting to compromise the set of machines on $ap_1$. In our computation, the NCP metric yields a value of 51.23\%. In the NCP metric the more machines are compromised, the higher the NCP value. Hence, the goal of the administrator is to reduce the NCP value. The vulnerable host percentage metric yield a value of 100\%. This is because all host in our example network has one vulnerability. This security metric is used to compute the percentage of host on a network that have at least one vulnerability.  

\section{Composite Security Metrics}
\label{NewMetrics}
\begin{figure}[hbt!]
	\centerline{\includegraphics[width=6in]{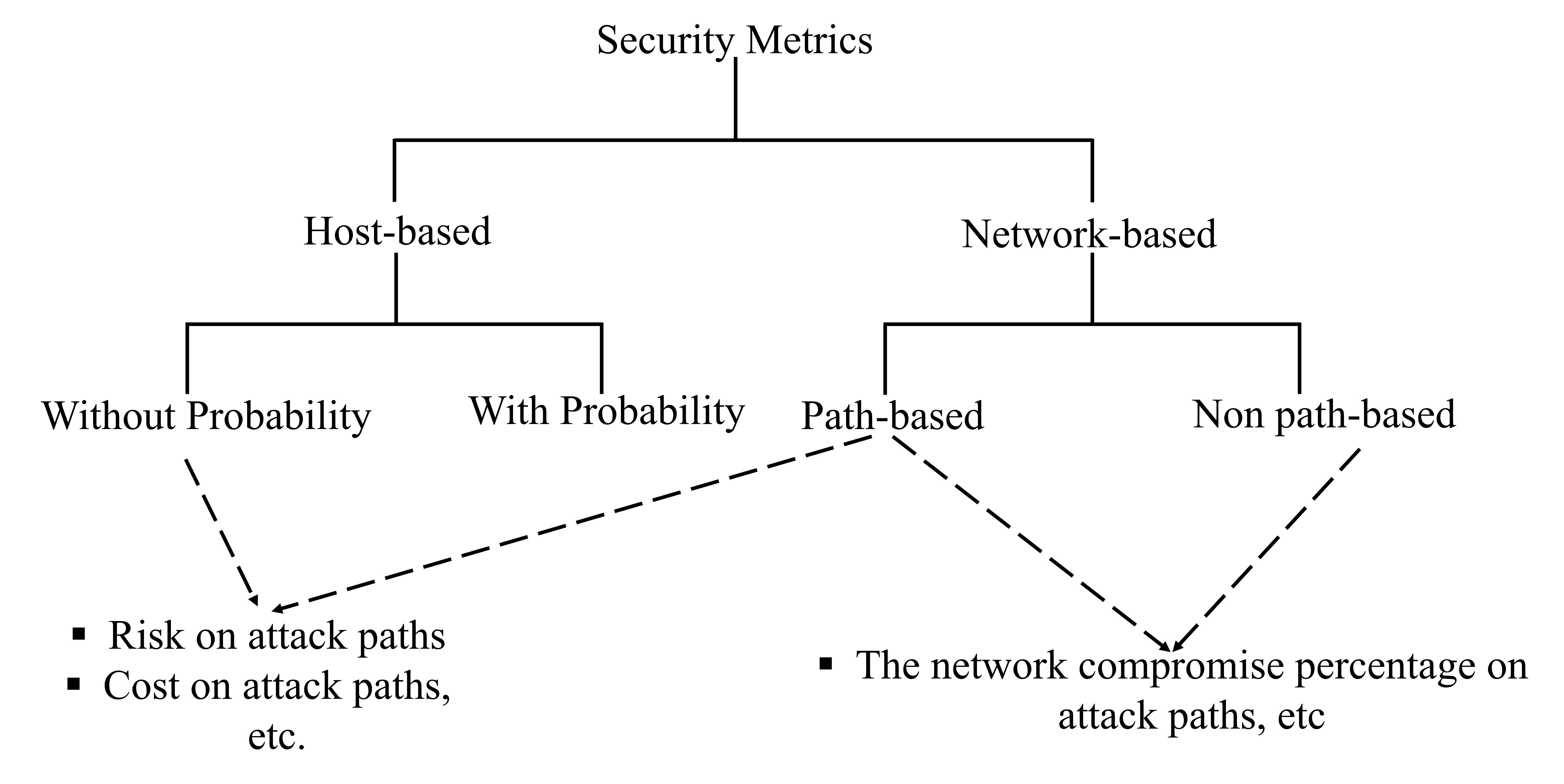}}
	\caption[]{Examples of composite security metrics.}%
	\label{fig_Composite}
\end{figure}

We propose an approach to develop new set of cyber security metrics called \textit{composite security metrics}. In these metrics, we combine individual metrics to create a new metric (for example, we can combine attack impact and attack path metric to form the impact on attack path metric, see Figure \ref{fig_Composite} for more examples). We will use the example network in Figure \ref{fig_Example} to perform security analysis using the composite security metrics. We demonstrate our proposed composite metrics using four examples: (i). Impact on attack paths (ii). Risk on attack paths (iii). Return on attack paths (iv). Probability of attack success on paths 

\subsection{Impact on Attack Paths}
The native metric (as one of the path-based metrics) used to create the impact of paths is attack paths. We combine the attack path metrics with the impact of each host in the path. We define the impact on attack path as the cumulative quantitative measure of potential harm in an attack path. We denote the metric as $AIM$ and calculate it using Equation (\ref{ImpactPath}) and (\ref{eq_AIM}). The host attack impact is calculated by equation (\ref{eq_aim_t2}). The network-level value $AIM$ is then given by Equation (\ref{eq_AIM}).

\begin{equation} \label{eq_aim_t1}
aim_{b}=
\left\{
\begin{array}{ll}
\sum\limits_{a \in c(b)}{aim_a}, & b \in B \atop g(b)=AND\\
\mathop {max} \limits_{a \in c(b)}{aim_a}, & b \in B \atop g(b)=OR
\end{array}
\right.
\end{equation}

\begin{equation} \label{eq_aim_t2}
aim_h=aim_{root}
\end{equation}

\begin{equation} \label{ImpactPath}
\begin{array}{ll}
aim_{ap}=\sum\limits_{h \in ap}{aim_h}, & ap\in AP
\end{array}
\end{equation}

\begin{equation} \label{eq_AIM}
AIM=\mathop {max} \limits_{ap \in AP}{aim_{ap}}
\end{equation}

The impact on path metric can reveal the impact of damage associated with each attack path. A security administrator can use this metric to determine which path to patch first. For instance, hosts in the path with the highest impact value can be considered as the prioritised set of hosts to patch.

Using the example network, we use all the possible AP from Figure \ref{fig_Example} to compute the impact of path metrics. 

\begin{align*} 
aim_{ap_1} &= aim_{h_1}+aim_{h_3}+aim_{h_4}+aim_{h_7}\\
&= 7+5+3+4.3 \\
&=19.3
\end{align*}
\begin{align*} 
aim_{ap_2}&= aim_{h_2}+aim_{h_5}+aim_{h_7}\\
&=4+9+4.3 \\
&=17.3
\end{align*}
\begin{align*} 
aim_{ap_3}&=aim_{h_2}+aim_{h_6}+aim_{h_7}\\
&=4+6.5+4.3 \\
&=14.8
\end{align*}

The $AIM$ of the example network is 19.3. More detail of how to get the CVSS impact values can be found in \cite{CVSSCalulator:2016}.\\

\subsection{Risk on Attack Paths}
The Risk on attack paths is defined as the expected value of the impact on an attack path. It is computed as the summation of the product of the probability of attack success \begin{math}pr_{h} \end{math} and the amount of damage \begin{math} aim_{h} \end{math} \textit{h} belonging to an attack path \textit{ap}. The metric is denoted as $R$ and calculate it using Equation (\ref{eq_maxrisk}). The host risk metric is defined by equation (\ref{eq_r2}). The network-level value $R$ is then given by Equation (\ref{eq_maxrisk}).

\begin{equation} \label{eq_r}
r_{b}=
\left\{
\begin{array}{ll}
\sum\limits_{a \in c(b)}{pr_a\times aim_a}, & b \in B \atop g(b)=AND\\
\mathop {max} \limits_{a \in c(b)}{pr_a\times aim_a}, & b \in B \atop g(b)=OR
\end{array}
\right.
\end{equation}

\begin{equation} \label{eq_r2}
r_h=r_{root}
\end{equation}

\begin{equation} \label{eq_aim_ap}
\begin{array}{ll}
r_{ap}=\sum\limits_{h \in ap}{pr_h \times aim_h}, & ap\in AP
\end{array}
\end{equation}

\begin{equation} \label{eq_maxrisk}
R=\mathop {max} \limits_{ap \in AP}{r_{ap}}
\end{equation}

We compute the risk of paths metric for all the possible attack paths as follows:
\begin{align*} 
r_{ap_1} &= pr_{h_1}\times aim_{h_1}+pr_{h_3}\times aim_{h_3}+ pr_{h_4}\times aim_{h_4}+pr_{h_7}\times aim_{h_7}\\
&= (0.75\times 7) + (0.43\times 5) + (0.21\times 3) + (0.44 \times 4.3) \\
&=9.92
\end{align*}
\begin{align*} 
r_{ap_2} &= pr_{h_2}\times aim_{h_2}+ pr_{h_5}\times aim_{h_5}+pr_{h_7}\times aim_{H_7}\\
&= (0.35\times 4) + (0.93\times 9) + (0.44 \times 4.3) \\
&=11.66
\end{align*} 
\begin{align*} 
r_{ap_3} &= pr_{h_2}\times aim_{h_2}+ pr_{h_6}\times aim_{h_6}+pr_{h_7}\times aim_{h_7} \\
&= (0.35\times 4) + (0.71\times 6.5) + (0.44 \times 4.3)\\
&=7.91
\end{align*}

This metric shows the level of risk associated with each attack path. From our computed example HARM, the attack path $ap_2$ (it's risk is 11.66) is considered as the path with the highest risk.

\subsection{Return on Attack Paths}
The return on attack \cite{Cremonini:ROA2005} is a metric used to quantify the benefit for the attacker. A return on attack paths computes the benefit for an attacker when the attacker successfully exploits all the vulnerabilities on a particular attack path. 
From the defender's point of view, the network administrator can use this metric to reduce the attacker's benefit by patching vulnerabilities on the path(s) with a high value of ROA. We denote the metric as $ROA$ and it is calculated using Equation (\ref{eq_maxroa}). The host return on attack metric is given by equation (\ref{eq_roaHost}). The network-level value $ROA$ is then given by Equation (\ref{eq_maxroa}).

\begin{equation} \label{eq_roa}
roa_{b}=
\left\{
\begin{array}{ll}
\sum\limits_{a \in c(b)}{\frac{pr_a\times aim_a}{ac_a}}, & b \in B \atop g(b)=AND\\
\mathop {max} \limits_{a \in c(b)}{\frac{pr_a\times aim_a }{ac_a}}, & b \in B \atop g(b)=OR
\end{array}
\right.
\end{equation}

\begin{equation} \label{eq_roaHost}
roa_h=roa_{root}
\end{equation}

\begin{equation} \label{eq_roa1}
\begin{array}{ll}
roa_{ap}=\sum\limits_{h \in ap}{\frac{pr_h\times aim_h }{ac_h}}, & ap\in AP
\end{array}
\end{equation}

\begin{equation} \label{eq_maxroa}
ROA=\mathop {max} \limits_{ap \in AP}{roa_{ap}}
\end{equation}

We show how to compute return on attack paths below:
\begin{align*} 
roa_{ap_1} &= \frac{pr_{h_1} \times aim_{h_1}}{ac_{h_1}} + \frac{pr_{h_3} \times aim_{h_3}}{ac_{h_3}} + \frac{pr_{h_4} \times aim_{h_4}}{ac_{h_4}} + \frac{pr_{h_7} \times aim_{h_7}}{ac_{h_7}} \\
&= \frac{0.25 \times 7}{8} + \frac{0.57 \times 5}{5} + \frac{0.79 \times 3}{3.5} + \frac{0.56 \times 4.3}{5.5}\\
&= 1.91
\end{align*}

\begin{align*} 
roa_{ap_2} &= \frac{pr_{h_2} \times aim_{h_2}}{ac_{h_2}} + \frac{pr_{h_5}\times aim_{h_5}}{ac_{h_5}} + \frac{pr_{h_7} \times aim_{h_7}}{ac_{h_7}}\\
&= \frac{0.65 \times 4}{4.2} + \frac{0.07 \times 9}{9.2} + \frac{0.56 \times 4.3}{5.5}\\
&= 1.12
\end{align*}

\begin{align*} 
roa_{ap_3} &= \frac{pr_{h_2} \times aim_{h_2}}{ac_{h_2}} + \frac{pr_{h_6}\times aim_{h_6}}{ac_{h_6}} +  \frac{pr_{h_7} \times aim_{h_7}}{ac_{h_7}}\\
&= \frac{0.65 \times 4}{4.2} + \frac{0.29 \times 6.5}{7.5} + \frac{0.56 \times 4.3}{5.5}\\
&= 1.30
\end{align*}

Return on attack paths quantifies the network security level from the attacker's perspective. From the example network scenario, the attack path $ap_1$ with metrics value 1.91 has the highest benefit to the attacker. 

\subsection{Probability of Attack Success on Paths}
The probability of attack success on paths is developed by combining path and probability of attack success. The probability of attack success on paths represents the chances of an attacker successfully reaching the target through an attack path. It is calculated by the equation (\ref{eq_pr}) 
The host attack success probability is defined by equation \ref{eq_asp_t2}. We denote probability of attack success on paths as $Pr$. The network-level value $Pr$ is then given by Equation (\ref{eq_pr}).

\begin{equation} \label{eq_asp_t1}
pr_{b}=
\left\{
\begin{array}{ll}
\prod\limits_{a \in c(b)}{pr_a}, & b \in B \atop g(b)=AND\\
1-\prod\limits_{a \in c(b)}(1-pr_a),  & b \in B \atop g(b)=OR
\end{array}
\right.
\end{equation}

\begin{equation} \label{eq_asp_t2}
pr_h=pr_{root}
\end{equation}

\begin{equation} \label{eq_asp_ap}
\begin{array}{ll}
pr_{ap}=\prod\limits_{h \in ap}{pr_h}, & ap\in AP
\end{array}
\end{equation}

\begin{equation} \label{eq_pr}
Pr=\mathop {max}\limits_{ap \in AP}{pr_{ap}}
\end{equation}

We show how to compute the probability of attack success on paths below: 
\begin{align*} 
pr_{ap_1} &= pr_{h_1}\times pr_{h_3} \times pr_{h_4} \times pr_{h_7} \\
&= 0.75 \times 0.43 \times 0.21 \times 0.44 \\
&= 0.03
\end{align*}
\begin{align*} 
pr_{ap_2} &= pr_{h_2}\times pr_{h_5} \times pr_{h_7}\\
&= 0.35\times 0.93 \times 0.44\\
&= 0.14
\end{align*}
\begin{align*} 
pr_{ap_3} &= pr_{h_2}\times pr_{h_6} \times pr_{h_7}\\
&= 0.35\times 0.71 \times 0.44\\
&= 0.11
\end{align*}

In this scenario, $ap_2$ with metric value 0.14 has the highest probability of a successful attack and therefore it is the $Pr$. The closer the $Pr$ value is to 1, the higher is the likelihood that an attacker will succeed in exploiting the target.  

\section{Conclusions and Future Work}
\label{conclusions}
In this paper, we have described the existing security metrics for cyber security assessment. We have used the network structure and reachability information to classify the existing metrics into host and network based security metrics. We also use the existing security metrics to carry out security analysis. In addition, we described an approach to developing composite security metrics and finally, we formally defined some composite security metrics.

Our classification of security metrics does not capture dynamic security metrics. Thus, we need to incorporate the dynamic security metrics into the proposed classification.

\bibliography{reference} 
\bibliographystyle{abbrv}

\end{document}